\documentstyle[11pt,aasms4,epsf]{article}

\def\folio{\ifnum\pageno<2\nopagenumbers\else\number\pageno\fi}
\newtoks\headline \headline={\hss\twelverm\folio\hss} 
\newtoks\footline \footline={{\hfil}} 
\font\mathbf=cmmib10 scaled 1000             

\def\ref{\par\noindent\hangindent=2pc \hangafter=1 }
\def\amin{\ifmmode^{\prime}\else$^{\prime}$\fi}
\def\asec{\ifmmode^{\prime\prime}\else$^{\prime\prime}$\fi}

\def\etal{{et al. }}
\def\cappage #1 #2 #3 {\vfill\eject\pageno=#1
\vglue 10 true in minus 10 true in \noindent{\bf Figure #2.} #3}
\def\ee #1 {\times 10^{#1}}
\def\ut #1 #2 { \, \hbox{#1}^{#2}}
\def\u #1 { \, \hbox{#1}}

\def\kms {\, \hbox{km}\,\hbox{s}^{-1}}

\def\percc {\, \hbox{cm}^{-3}}

\let\grad=\nabla
\def\cross{{\bf \times}}
\def\curl #1 {\grad \cross #1}
\def\div #1 {\grad \cdot #1}

\def\kms    {\hbox{km{\hskip0.1em}s$^{-1}$}}    
\def\etal   {{\it et al. }}                     

\begin{document}

\title{OH (1720 MHz) Masers and Mixed-Morphology Supernova Remnants}

\author{F. Yusef-Zadeh}
\affil{Department of Physics and Astronomy, Northwestern University,
Evanston, Il. 60208 (zadeh@northwestern.edu)}

\author{M. Wardle}
\affil{Department of Physics, Macquarie University, NSW 2109, Australia
(wardle@physics.mq.edu.au)}

\author{J. Rho}
\affil{SIRTF Science Center, California Institute of Technology, MS 220-6,
CA 91125
 (rho@ipac.caltech.edu)}

\author{M. Sakano}
\affil{Department of Physics and Astronomy,
University of Leicester, Leicester LE1 7RH, UK 
 (mas@star.le.ac.uk)}

\begin{abstract}

Radio surveys of supernova remnants (SNRs) in the Galaxy have
uncovered 19 SNRs accompanied by OH maser emission at 1720 MHz.  This
unusual class of maser sources is suggested  to be produced behind a
shock front from the expansion of a supernova remnant running into a
molecular cloud.  An important ingredient of this model is that
X-ray emission from the remnant enhances the production of OH
molecule.  The role  of X-ray emission from maser emitting (ME)  SNRs
is investigated by comparing  the X-ray induced ionization rate 
with theory.  
One aspect of this model is verified: there is a
strong association between maser emitting  and mixed-morphology (MM) or
thermal composite SNRs
--center-filled thermal X-ray emission surrounded by shell-like radio
morphology. We also present ROSAT and ASCA 
observations of  two maser
emitting SNRs: G21.8--0.6 (Kes 69) and G357.7--0.1 (Tornado).

\end{abstract}


\section{Introduction}

The interaction between supernova remnants and molecular clouds
constitutes an important part
of Galactic ecology. Such interactions should be common because 
massive stars do not drift far from their parent cloud 
during their short lifetime. Nearby supernova explosions are responsible 
for driving shocks
into molecular clouds as they heat,  stir, disrupt, change the 
chemical
evolution of the cloud and possibly trigger star formation. 

OH masers at 1720 MHz have recently provided powerful signatures 
of SNR-molecular cloud interaction sites
(Frail, Goss \& Slysh 1994; Wardle \& Yusef-Zadeh 2002). 
The masers arise in dense ($\sim 10^5 \percc$) gas with a
temperature in the range 50--125$\,$K -- conditions that occur in
shocked molecular gas (Elitzur 1976; Lockett, Gauthier \& Elitzur
1999).  The necessary OH abundance in the shocked gas
(OH/H$_2\ga 10^{-6}$) is created
by the dissociation of shock-produced water induced by thermal
X-rays emitted from the hot gas filling the adjacent supernova remnant
(Wardle 1999).
Nineteen remnants (about 1 in 10) have 1720 MHz masers,
and are therefore likely to be interacting with clouds (Frail \etal
1996; Green \etal 1997; Koralesky \etal 1998; Yusef-Zadeh \etal 1999).
The inferred interactions have been confirmed by follow up searches 
for millimeter or infrared emission from hot molecular gas
or from molecules produced by the rich chemistry occurring
within the shock front (e.g.  Reach \& Rho 1998; Frail \& Mitchell
1998; Reynoso \& Mangum 2000; Yusef-Zadeh \etal 2001; Lazendic \etal
2002a).

A supernova remnant's X-ray appearance is also believed to be affected by 
interaction with a molecular cloud or dense atomic gas (Rho \& Petre 1998).
The X-ray morphology of supernova remnants was originally divided into
shell-like, Crab-like (plerionic), and 
shell-like remnants containing plerions (composite) morphologies 
(Seward 1985).  An
additional ``mixed-morphology'' (MM)  class (sometimes called thermal
composites)  contains
approximately 25\% of X-ray detected remnants, which appear
center-filled in X-rays and shell-like at radio wavelengths (Rho 1995;
Rho \& Petre 1998).  These remnants  possess 
a soft thermal bremsstrahlung continuum and H-like and He-like metal 
lines with roughly solar abundances, characteristic
of hot interstellar gas rather than shocked ejecta
(Rho \& Petre 1996).
The interstellar gas in the interior of MM SNRs is accounted for
by the passage of a supernova shock propagating into dense and possibly
clumpy interstellar gas.  In one model, soft thermal X-ray emission
arises via the evaporation of clumps overrun by the supernova shock
(White and Long 1991).  Alternatively, heat
conduction within a remnant expanding into a moderate-density medium 
reduces the internal temperature and density gradients and is responsible for their X-ray appearance
(Chevalier 1999; Cui and Cox 1992; Cox et al.  1999; Shelton
et al.  1999).

There are reasons to expect significant overlap of the
mixed-morphology and maser-emitting classes of supernova remnants.  
Both are believed to arise through interaction with an adjacent
molecular cloud (although MM remnants may also arise through
expansion into a dense atomic region ), and the soft thermal X-rays
emitted from a mixed-morphology remnant appear to be a necessary
ingredient for enhancing OH behind the shock front (Wardle 1999). It
has been noted previously that several 1720 MHz remnants fall into
the mixed-morphology class (Green \etal 1997).  

Here we address the following two new key points related to the
nature of MM and ME SNRs: First, we address the physical 
relationship between two different class 
of objects,
MM and ME SNRs.  Although a number of papers have previously
suggested that some of the brightest and most well-known MM and ME
SNRs (e.g. W28, W44 and IC 443) are physically correlated, in none
of the previous studies, has an unbiased quantitative and statistical
analysis of all the ME SNRs  been made. Given the speculative
state of this correlation in previous studies, the present analysis
puts correlation of these two class of remnants on a strong footing
and shows that there is statistical evidence for an association
between these two classes of remnants. This correlation may imply
that MM remnants require to have dense molecular gas in their
environment to explain  their unusual centrally 
filled
X-ray morphology.

Second point we address here is to test the model by Wardle (1999) 
who
predicted that X-ray emission is one of the main ingredients
responsible for production of OH molecule behind a C-type shock.  In
order to test this model, we searched for X-ray emission from all
known SNR masers and measured X-ray ionization rate at the edge of
SNR masers whose thermal spectrum and whose MM  have previously 
been identified. 
We also report the first evidence of  X-ray detection 
from the Tornado nebula and Kes 69. 
 It should be pointed out that such an analysis to determine X-ray
ionization rate associated with SNR masers have never been made
previously.  Our analysis gives a strong support to the mechanism
responsible for production of OH masers with the implication that
shock chemistry incorporating X-ray induced ionization and 
dissociation   is important. 


\section{An Association between Mixed-morphology and Maser-emitting SNRs}

We have examined the X-ray properties of all known ME SNRs.  Among
190 SNRs that have been searched for OH (1720 MHz) maser emission,
19 detections are found in the disk and in the Galactic center (e.g.  
Green 1997; Yusef-Zadeh \etal 1999).  Table 1 shows the names of all
known 19 ME SNRs and their corresponding X-ray characteristics
compiled from observations using ASCA, ROSAT and {\it Einstein}
data.  The X-ray and radio properties of each remnant are taken from
references indicated in column ten.  These ME remnants fall into
three broad groups.  The first group of seven have MM counterparts,
the second group of eight show X-ray emission but future
observations are required to determine the morphology and the
spectrum of these remnants. Both groups are denoted Y in column 6.  
The remaining four remnants have unknown X-ray properties, either
being unobserved in X-rays or their soft X-ray emission have been
absorbed by intervening interstellar gas.

We note that the detection of X-ray emission from G357.7+0.1 (source 13), 
G21.8-0.6
(Kes 69) and G349.7+0.2 (Slane et al. 2002)
 in the second group of remnants are found from
 ASCA and ROSAT archival data.  Examining GIS
detector of ASCA's archival data, a weak X-ray emission at a level of
3--3.5$\sigma$  has been detected from the head of the Tornado nebula near
$\alpha = 17^h 40^m 10.1^s, \delta = -30^0 58' 8.4''$ (J2000). This
position coincides within 30$''$ of the compact maser position associated
with the Tornado nebula (Frail et al. 1996). The other remnant in the
second group in Table 1 is Kes 69 which has an incomplete radio shell
morphology and X-ray emission has been reported from Einstein observations
(Seward 1990).  Figure 1 shows contours of X-ray emission based on  ROSAT PSPC
observations superimposed on a grayscale NVSS radio continuum image at 20cm
(Condon et
al. 1998). The brightest X-ray feature lies in the interior of the 
radio shell surrounded by a number of X-ray blobs forming two halves of a
shell-like structure.  An X-ray feature is also noted to the east possibly
coincident with the radio shell near $\alpha = 18^h 33^m 18^s, \delta =
-10^0 09'$ (J2000).  Many of the radio and X-ray features are distributed 
to
the south
delineating a partial arc. The northern half of the
radio shell is weak and is best detected in  low-resolution 90cm image of Kes 69
by Kassim (1992). The cross shows the position of the OH(1720 MHz) maser
associated with the remnant (Green et al. 1997). 

We have extracted PSPC
spectra from the entire SNR G21.8-0.6 (Kes 69) excluding the bright 
point source near 
$\alpha = 18^h 32^m 50.677^s, \delta = -10^0 01' 15''$ (J2000)
and fit thermal and non-thermal models. Both
models yield reduced $\chi^2$ of 1.1, but the power law index in the
non-thermal fit is negative, which is non-physical, maybe suggesting the
emission is thermal. The best fit is using a thermal model 
(Mewe,  Groneschild and ven den Oord 1985;  a.k.a. ''vmekal'' in XSPEC)
with N$_H$ = 2.4$\times 10^{22}$ cm$^{-2}$ and kT = 1.6 keV. The inferred
luminosity is 3.5$\times 10^{35}$ erg s$^{-1}$ assuming the distance of
11.2 kpc. Assuming a distance of 11.2 kpcs and a geometrical mean angular
size of $\approx10'$, estimated from the distribution of X-ray image in
Figure 1, the X-ray ionization rate ($\zeta$) is estimated to be $\approx
10^{-16}$ s$^{-1}$. However, there are
considerable uncertainties in making this estimate including the angular
size and the distribution of the X-ray emission with respect to radio
emission. Other uncertainties include the distance determination ranging
between 6.3 and 11.2 kpc and the spectral model due to poor spectral
resolution of PSPC and the presence of multi-temperature components.  Thus,
the bright X-ray emission from the interior of the radio shell and a
possible thermal X-ray spectrum suggest that this remnant may be a
mixed-morphology SNR, but due to the poor quality of the data, future
observations are required to determine the nature of this source.
Similarly, better distance determination as well as X-ray morphology are
needed not only for G21.8-.06 but also G349.7+0.2 in order to include them
in group 1.
Slane et al. (2002) 
have recently reported X-ray emission from G349.7+0.2 
and estimated X-ray luminosity of
1.8$\times10^{37}$ erg s$^{-1}$ at a 
distance of 22 kpc. Using the angular size of 1$'$ (6 pc),  
the X-ray ionization
rate  is estimated to be $\sim2\times10^{-14}$.

At present there are no counterexamples to the conjecture that
all SNRs with OH(1720 MHz) masers are mixed-morphology.  Even if all
of the  SNRs with unknown X-ray morphology are assumed to {\it{not}} be
mixed-morphology, the number of mixed-morphology remnants with 1720
MHz masers (i.e.
7) is much higher than random given that 10\% of SNRs are ME
(Koralesky \etal 1998) and that 7\% of SNRs are MM (Rho \& Petre 1998).
We can quantify this by using a contingency table (Sokal \& Rolfe
1995; see also Press et al. 1992)  to test the null hypothesis that there
is no association between MM and ME remnants (see Table 2).  The bold
entries in Table 2 give the number of SNRs broken down by presence or
absence of OH(1720 MHz) masers and their X-ray properties (i.e.  mixed
morphology, not mixed morphology and unknown), giving six categories
of remnant.  The numbers for SNRs
with 1720 MHz masers in the first row are derived from Table 1.  The
numbers for SNRs {\it{without}} 1720 MHz masers listed (in bold) in the
second row of Table 2 are less certain,
being based on the total of 15 MM remnants identified by Rho \& Petre
(1998) and their estimate that this represents about 25\% of X-ray
detected SNRs.  However, our results are not sensitive to these
numbers, nor to the fact that only 190 of the 225 known galactic SNRs
have been surveyed for 1720 MHz masers.  The second entry in each of
the six SNR categories is the expected number of remnants derived from
the counts assuming that the maser and X-ray characteristics are
completely independent of one another.  Finally, the third entry gives
the contribution to $\chi^2$, (observed-expected)$^2/$(expected) from
each category. 
The total $\chi^2$ is 27.5 with 2 degrees of freedom, indicating that the
hypothesis of independence can be rejected at the level of
$1.1\times 10^{-6}$, in other words that the two classes are associated.
 Most
of the contribution to $\chi^2$ comes from the 7 remnants in the first
group - by chance one would expect 1.5.  We have, of course, assumed
that there is no selection bias in the observations.  The most obvious
bias - that X-rays observations are limited by interstellar
absorption, whereas radio observations are not, does not cause obvious
problems.

\section{X-ray Shock Chemistry}

We have also examined whether the X-ray emission from the individual
remnants in the first group of Table 1 is sufficient to dissociate
water molecules behind a C-type shock front and produce the OH column
necessary for OH(17200 MHz) masers.  The X-ray induced ionization rate
in the gas should be $\zeta \ga 10^{-16}\ut s -1 $ (Wardle 1999).  The
X-ray ionization rate at the edge of the remnant, listed in column 9,
is estimated from the total X-ray luminosity of individual remnants
using $\zeta = N_e \sigma F_X$, where $N_e = 30\ut keV -1 $ is the
mean number of primary and secondary electrons generated per unit
energy deposited by X-rays, $\sigma \approx 2.6\ee -22 \ut cm 2 $ is
the photoabsorption cross section per hydrogen nucleus at 1 keV, and
$F_X \approx L_X/4\pi r^2$ is the X-ray intensity at the edge of the
remnant.  The values of the ionization rate ($\zeta$) of ME SNRs with
MM counterparts is consistent with the predicted range within a factor
of few.

Significant uncertainty arises from the asymmetric and clumpy
distribution of the X-ray emission, differential extinction across the
face of the remnant and the estimated column density of neutral gas
along the line of sight based on fitted spectrum of the remnant.  One
example of this uncertainty is as follows.  Two studies of G359.1-0.5
based on ASCA and ROSAT give different values of hydrogen column and
soft X-ray flux (Egger and Sun 1998; Bamba et al.  2000).  The
unabsorbed X-ray flux
differs by a factor of 25 due to different thermal plasma models used
by these authors.  Both studies give low value of X-ray flux at the
edge of the remnant, F$_x$ in column 8.  We selected the higher flux
values given by Egger and Sun (1998) because G359.1-0.5 is surrounded
by two bright hard X-ray sources and hard X-ray emission from the
Galactic center region.  Source confusion may contaminate the
background subtraction, particularly the low-resolution ASCA data
despite the careful analysis of Bamba et al. (2000).  G359.1-0.5
is a highly symmetric shell-type SNR at radio wavelengths whereas the
X-ray emission is concentrated mostly to one side of the interior.
Reducing the size of X-ray emission by a factor of two increase F$_x$
by a factor of four which is consistent with the predicted value of
3$\times10^{-16}$ s$^{-1}$ (Wardle 1999).

\section{Implications}

The strong association between MM and ME remnants plus the fact ME
remnants are interacting with molecular clouds suggest that many 
MM
remnants are created by the interaction with a dense cloud or 
the denser-than average environment surrounding molecular 
clouds (White
and Long 1991; Chevalier 1999; Shelton et al. 1999).  The MM 
remnants without OH(1720
MHz) masers may be formed as a result of expansion into a moderate
density medium (Shelton et al 1999).  Alternatively, all MM remnants
could arise from interaction with molecular clouds, with masers
failing to arise because of the restricted physical conditions under
which compact maser emission is formed (Lockett \etal 1999) or 
the masers could simply be beamed away from the line of sight.  These
effects could explain the lack of ME counterparts for eight of the
MM SNRs, tabulated by Rho \etal (1998).  It would be useful to
examine the eight MM remnants without maser emission for other, more
generic, signatures of molecular cloud interactions such as OH 
absorption or emission from hot H$_2$
at 2 $\mu$m.

Within the uncertainty of our heterogeneous samples, the 
estimated  X-ray-induced ionization rates are sufficient to
enhance the OH behind shock waves driven into molecular clouds by SNRs
(Wardle 1999).  It appears that the ionization rates in molecular
clouds adjacent to SNRs, $\sim 3\ee -16 \ut s -1 $, are 10--30 times
greater than generally adopted.  It would be interesting to consider
the consequences for chemistry within the clouds. In particular, 
future  millimeter, submillimeter and infrared observations 
should be useful to determine  the role of X-ray shock 
chemistry in the ME and MM SNRs.

Acknowledgments: We are grateful to Mike Wheatland for suggesting the
contingency
table analysis and to Jasmina  Lazendic for discussing her results 
prior
to publication. We also thank the referee for useful comments. 
MS and FYZ acknowledge partial support from Japan
Society of the Promotion of Science and NASA, respectively.

\vfill\eject

 \begin{figure}
\figcaption{Contours of X-ray emission based on ROSAT PSPC observations 
are superimposed on a grayscale radio continuum 
image of Kes 69 based on NVSS observations at 20cm. 
The cross coincides with the postion of OH(1720 MHz) maser at v$_{LSR} = 69.3$
\kms (Green et al. 1997).}
 \end{figure}


\begin{references}



\reference{} Asaoka, I. \& Aschenbach, B. 1994 A\&A, 284, 573

\reference{} Bamba, A. Yokogawa, J., Sakano, M. \& Koyama, K.  2000,
PASJ, 52, 259

\reference{} Chevalier, R.A.  1999, ApJ,  511, 798


\reference{c97} Claussen, M.J., Frail, D.A., Goss, W.M., \& Gaume,
R.A. 1997, ApJ, 489, 143

\reference{b98} 
Condon, J. J.,  Cotton, W. D., Greisen, E. W.,  Yin, Q. F., 
Perley, R. A., Taylor, G. B. \& Broderick, J. J. 1998, AJ, 115, 1693


\reference{}  Cox, D.P., Shelton, R.L., Maciejwski, W., Smith, R., Plewa,
T. et al. 1999,  ApJ,  524, 179

\reference{} Cui, W. \& Cox, D.P.   1992, ApJ,  178, 159



\reference{e76} Elitzur, M. 1976, ApJ, 203, 124

\reference{e76} Egger, R. \& Sun, X.  1998, in the Proc. of IAU Colloq.
No.
166, eds: Brietschwerdt, M.J., Freyberg, J. and J. Trumper, Lecture Notes
in  Physics 506 (Springler-Verlag, Berlin) p417



\reference{f96} Frail, D. A., Goss, W. M., Reynoso, E. M.,
Giacani, E.B., Green, A. J. \& Otrupcek, R. 1996, AJ, 111, 1651

\reference{fgs94} Frail, D.A., Goss, M.W. \& Slysh, V.I. 1994, ApJ,
  424, L111


\reference{g114} Green A.J., Frail, D.A., Goss, W.M. \& Otrupcek, R.
 1997, AJ, 114, 2058




\reference{k98} Kassim, N. 1992, AJ, 103, 943

\reference{} Koo, B.-C., Kim, K.T., \& Seward, F.D. 1995, ApJ, 447,
211

\reference{k98} Koralesky, B., Frail, D.A., Goss, W.M.,
Claussen, M.J. \& Green, A.J. 1998, AJ, 116, 1323

\reference{} Lazendic, J.S. et al. 2002b, in preparation


\reference{} Lazendic, J.S.,
Wardle, M., Burton, M.G., Yusef-Zadeh, F., Whiteoak, J.B.,
Green, A.J.  \etal 2002a, MNRAS, 331, 537


\reference{l89} Leahy, D.A. 1989, A\&A, 216, 193

\reference{l98} Lockett, P., Gauthier, E. \&  Elitzur, M. 1998,
ApJ, 511, 235


\reference{} Maeda, Y.,  Baganoff, F. K.
Feigelson, E. D.,  Morris, M., Bautz, M. W. et al.  2002, ApJ, 570, 671

\reference{} Mewe, R., Groneschild, E.H.B.M., \& ven den Oord, G.H.J. 
1985, A\&AS, 62, 197


\reference{} Press, W. \etal 1992, Numerical Recipes in Fortran
 (Cambridge University Press: Cambridge)

\reference{} Reach, W. T. \& Rho, J. 1998, ApJ, 507, L93

\reference{} Reynoso, E.M. \& Mangum, J.G.   1999, ApJ,  511, 798

\reference{}
Rho, J. 1995, An X-ray study of composite supernova remnants,
PhD thesis, University of Maryland

\reference{}
Rho, J. \& Berkowski, K.J.   2002, ApJ (in press)

\reference{}
Rho, J. \& Petre, R. 1996, ApJ, 467, 698

\reference{}
Rho, J. \& Petre, R. 1998, ApJ, 503, L167


\reference{} Rho, J.  Petre, R.  Schlegel, E.M. \& Hester, J.J. 1994,
ApJ, 430, 757



\reference{}
Seward, F.D. 1985, Comments Astrophy. XI, 1, 15


\reference{} 
Seward, F.D. 1990, ApJS, 73, 781

\reference{}
Sidoli, L., Merghetti, S., Treves, A., Parmer, Turolla, R. et al. 
2001, A.\&A., 372, 651

\reference{} Shelton, R.L., Donald. P. Cox, D.P., Maciejewski, Smith, W.
et al.   1999, ApJ,  524, 192

\reference{} Slane, P., Chen, Y., Lazendic, J. \& Hughes, J. 2002, 
ApJ (in press). 



\reference{} Sokal, R. R. \& Rohlf, F. J. 1995,
Biometry : the principles and practice of statistics in biological research
(New York: W.H. Freeman)

\reference{}
Sugizaki, M., Mitsuda, K., Kaneda, H., Matsuzaki, K., Yamauchi, S. et al.
2001, 134, 77

\reference{}
Wardle, M.  1999, ApJ, 525, L101.

\reference{}
Wardle, M. \& Yusef-Zadeh, F. 2002, Science, 296, 2350

\reference{} White, R.L. \& Long, K.S. 1991, ApJ, 373, 543


\reference{y95} Yusef-Zadeh, F., Uchida, K.I., \& Roberts, D.A. 1995,
  Science 270, 1801

\reference{y96} Yusef-Zadeh, F., Roberts, D.A., Goss, W.M., Frail,
D.A. \&
  Green, A. 1996, ApJ 466, L25

\reference{y99} Yusef-Zadeh, F., Roberts, D.A., Goss, W.M., Frail,
D.A. \& Green, A. 1999, ApJ, 512, 230.



\reference{}
Yusef-Zadeh, F.,  Stolovy, S. R., Burton, M., Wardle, M., Ashley, M. C. B.
2001, ApJ, 560, 749



\reference{}


\end{references}
\end{document}